\documentclass{article}

\usepackage{arxiv}
\usepackage{amsmath,amssymb,amsfonts}

\usepackage[utf8]{inputenc} % allow utf-8 input
\usepackage[T1]{fontenc}    % use 8-bit T1 fonts
\usepackage{hyperref}       % hyperlinks
\usepackage{url}            % simple URL typesetting
\usepackage{booktabs}       % professional-quality tables
\usepackage{amsfonts}       % blackboard math symbols
\usepackage{nicefrac}       % compact symbols for 1/2, etc.
\usepackage{microtype}      % microtypography
\usepackage{cleveref}       % smart cross-referencing
\usepackage{lipsum}         % Can be removed after putting your text content
\usepackage{graphicx}
\usepackage{doi}

\usepackage{algorithmic}
\usepackage{graphicx}
\usepackage{textcomp}
\usepackage{xcolor}

\usepackage[nolist,nohyperlinks]{acronym}
\usepackage{xurl}
\usepackage{comment}
\def\BibTeX{{\rm B\kern-.05em{\sc i\kern-.025em b}\kern-.08em
    T\kern-.1667em\lower.7ex\hbox{E}\kern-.125emX}}

\usepackage{tikz}
\usetikzlibrary{arrows,positioning,shapes.geometric, trees}

\title{On the Security of IO-Link Wireless Communication in the Safety Domain}

\author{ \href{https://orcid.org/0000-0002-6882-1214
}{\includegraphics[scale=0.06]{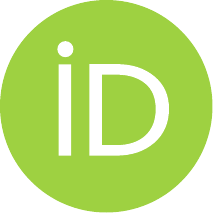}\hspace{1mm}Thomas R.~Doebbert}
\thanks{© 2022 IEEE.  Personal use of this material is permitted.  Permission from IEEE must be obtained for all other uses, in any current or future media, including reprinting/republishing this material for advertising or promotional purposes, creating new collective works, for resale or redistribution to servers or lists, or reuse of any copyrighted component of this work in other works.} \\
	Department of Electrical Measurement Engineering\\
	Helmut-Schmidt-University\\
	Hamburg, Germany \\
	\texttt{thomas.doebbert@hsu-hh.de} \\
	\And
	\href{https://orcid.org/0000-0002-4532-731X}{\includegraphics[scale=0.06]{orcid.pdf}\hspace{1mm}Florian Fischer} \\
	HSA\_innos\\
	University of Applied Sciences Augsburg\\
	Augsburg, Germany  \\
	\texttt{florian.fischer@hs-augsburg.de} \\
	\AND
	Dominik Merli \\
	HSA\_innos\\
	University of Applied Sciences Augsburg\\
	Augsburg, Germany  \\
	\texttt{dominik.merli@hs-augsburg.de}\\
	\And
	Gerd Scholl\\
	Electrical Measurement Engineering\\
	Helmut-Schmidt-University\\
	Hamburg, Germany \\
	\texttt{gerd.scholl@hsu-hh.de}\\
}

\renewcommand{\shorttitle}{On the Security of IO-Link Wireless Communication in the Safety Domain}

\hypersetup{
pdftitle={A template for the arxiv style},
pdfsubject={q-bio.NC, q-bio.QM},
pdfauthor={David S.~Hippocampus, Elias D.~Striatum},
pdfkeywords={First keyword, Second keyword, More},
}

\begin{document}
\maketitle

\begin{abstract}
  Security is an essential requirement of \ac{ICS} environments and its underlying communication infrastructure.
  %TODO within?
  Especially the lowest communication level within \ac{SCADA} systems - the field level - commonly lacks security measures.

  Since emerging wireless technologies within field level expose the lowest communication infrastructure towards potential attackers, additional security measures above the prevalent concept of air-gapped communication must be considered.

  Therefore, this work analyzes security aspects for the wireless communication protocol \ac{IOLW}, which is commonly used for sensor and actuator field level communication.
  A possible architecture for an IOLW safety layer has already been presented recently \cite{doebbert_cammin_scholl_2022}.

%  in safety applications.

  In this paper, the overall attack surface of \ac{IOLW} within its typical environment is analyzed and attack preconditions are investigated to assess the effectiveness of different security measures. Additionally, enhanced security measures are evaluated for the communication systems and the results are summarized.
  Also, interference of security measures and functional safety principles within the communication are investigated, which do not necessarily complement one another but may also have contradictory requirements.

% TODO mentioned, that safety security combination can lead to contradictions
 This work is intended to discuss and propose enhancements of the \ac{IOLW} standard with additional security considerations in future implementations.
\end{abstract}

% keywords can be removed


\acresetall

\section{Introduction}

% Story: Overall story
Due to the steadily growing trend of interconnections within \ac{ICS} and \ac{CPS} in Industry 4.0, also wireless technologies are increasingly employed across all communication domains.

%Interconnection and wireless technology emerge drastically within \ac{ICS} due to the trend/processing of Industry 4.0.

This connectivity pervades all communication layers within typical \ac{SCADA} systems from management and planning level to the lower control and field level.
Therefore, the attack surface increases not only within \ac{IP} based networks but also within the lowest field level.

Air-gapping field level systems is considered a sufficient protection measure in combination with physical access control.
Especially the wireless technology can enlarge the attack surface drastically within this low level of communication.

But even if field level wireless communication is surely not the most targeted part for attacks on \ac{ICS} systems, reasonable security measures should be implemented to not become the weakest link within a holistic security concept.

Future areas of applications for field level communication may be found in domains (e.g. mobile roaming device) where physical segregation are unfeasible, therefore security aspects become even more urgent.
% TODO Agricultural processes, public transport, mobile safety applications

% IOLW
\ac{IOLW} is an example for a wireless field level communication protocol, commonly used below the field bus level. %TODO 1-3 Beispiele nennen? % and maybe in the future for safety applications.
As a safety specification already exists for the wired IO-Link standard \cite{IOLSafetySysExtensions}, a safety extension for the wireless equivalent shall be feasible in the future.

In the first phase of the \ac{IOLW} standardization process, security risks considerations were not the focus of the protocol design and therefore, at the moment, no security measures are part of the protocol specification \cite{iolw}.
To protect the \ac{IOLW} communication within future use cases, security measures are necessary and must be defined and implemented.

While devices that use \ac{IOLW} communication protocol are constrained in its resources, common state of the art asymmetric cryptographic methods, e.g. \ac{TLS}, are not feasible here.
Thus, trade offs must be accepted, while increasing the security of \ac{IOLW} communication by security measures.
A recent proposal already outlined practical security measures for \ac{IOLW}, while orienting on common state of the art wireless protocols and its security measures \cite{doebbert_cammin_scholl_2022}.

% Description what's the targeted contribution

The contribution of this work is to determine the overall risk for \ac{IOLW} communication systems, therefore relevant attack scenarios are depicted and the impact from safety and security perspective is investigated.
Moreover, already proposed security measures, which are intended to protect the communication,  \cite{doebbert_cammin_scholl_2022} are analyzed.
Further security enhancements are derived and its applicability is investigated.

%paper structure as follows
The paper starts with a brief background of \ac{IOLW} with its key aspects in Section \ref{sec:IOLW}.
In Section \ref{sec:Methodology} the methodology of the security analysis is depicted.
This analysis is described in Section \ref{sec:SecAnalysis}.
Furthermore, in Section \ref{sec:InfluenceSecuritySafety} influences of security measures on safety are outlined.
The conclusion in Section \ref{sec:conclusion}, findings are summarized and future work is discussed.

\section{IO-Link Wireless}
\label{sec:IOLW}
%================================================================================================

\ac{IOLW}  was developed as an extension of the proven IO-Link standard \cite{ILC2019}, which is known as Single-Drop digital communication interface (SDCI) or IEC 61131-9 \cite{IEC2013cc}.
Within the factory automation structure, IOLW is mainly intended for sensor/actuator field-bus communication \cite{ILC2019, IEC2013cc, Heynicke2018}.

%TODO vielleicht kürzen
General surveys of IOLW as an open-vendor communication solution for factory automation on the shop floor are given in \cite{Heynicke2017,Heynicke2018,Wolberg2018,Rentschler2018, DS2CCP} with a focus on roaming in \cite{Rentschler2017}, antenna planning in \cite{Cammin2019a}, coexistence in \cite{coexistence, Solzbacher01Sep.2018, IOLWcoexTool}, security enhancement in \cite{IOLWcrypto, IOLWcryptoPrecisionSMSI2021} and functional safety \cite{doebbert_cammin_scholl_2022}, and on IOLW testing in \cite{cammin-jsss2018, Cammin2019, Cammin2019b, Cammin2020, Cammin2021}. Nevertheless, a short introduction to IOLW is given here.

In a star-shaped topology, IOLW offers bidirectional wireless communication for (cyclic) process data and (acyclic) on-request data between a wireless master (``W-Master'') and wireless devices (``W-Devices'') \cite{iolw, Heynicke2018}.
The 2.4 GHz ISM band with gaussian frequency shift keying (GFSK) modulation based on the physical layer (PL) of Bluetooth Low Energy (BLE) 4.2 is utilized in combination with a frequency- and time division multiple access (F/TDMA) scheme and a frequency hopping table.
The wireless coexistence behavior can be improved by omitting single frequency channels using blocklisting \cite{Heynicke2018, IOLWcoexTool}.
Wireless bridges (W-Bridges) are standardized to behave similar to W-Devices while offering a wired IO-Link port in order to retrofit legacy systems.
 %spaeter wichtig?

Up to three W-Masters can operate in the same manufacturing cell with each W-Master providing one to five tracks and each track supporting up to eight slots. Single-slot (SSlot) and double-slot (DSlot) W-Devices are specified.
SSlot W-Devices offer two (one) octet(s) for payload and are intended for simple sensors or actuators such as switches.
DSlot W-Devices offer 15 (14) octets for payload and are suitable for smart sensor applications (the values in the parenthesis include or exclude the obligatory control octet).
Within one track, SSlot and DSlot W-Devices can be combined \cite{iolw}. Up to 120 (SSlot) W-Devices are supported within e.g. a single manufacturing cell \cite{iolw, Heynicke2018}. Furthermore, \ac{IOLW} includes mechanisms to support energy-self-sufficient W-Devices and roaming of W-Devices or W-Bridges between different cells \cite{Rentschler2017}. %TODO fehlt in literatur?

\begin{figure}[h!]
	\centering
	\includegraphics[width = 0.45\textwidth]{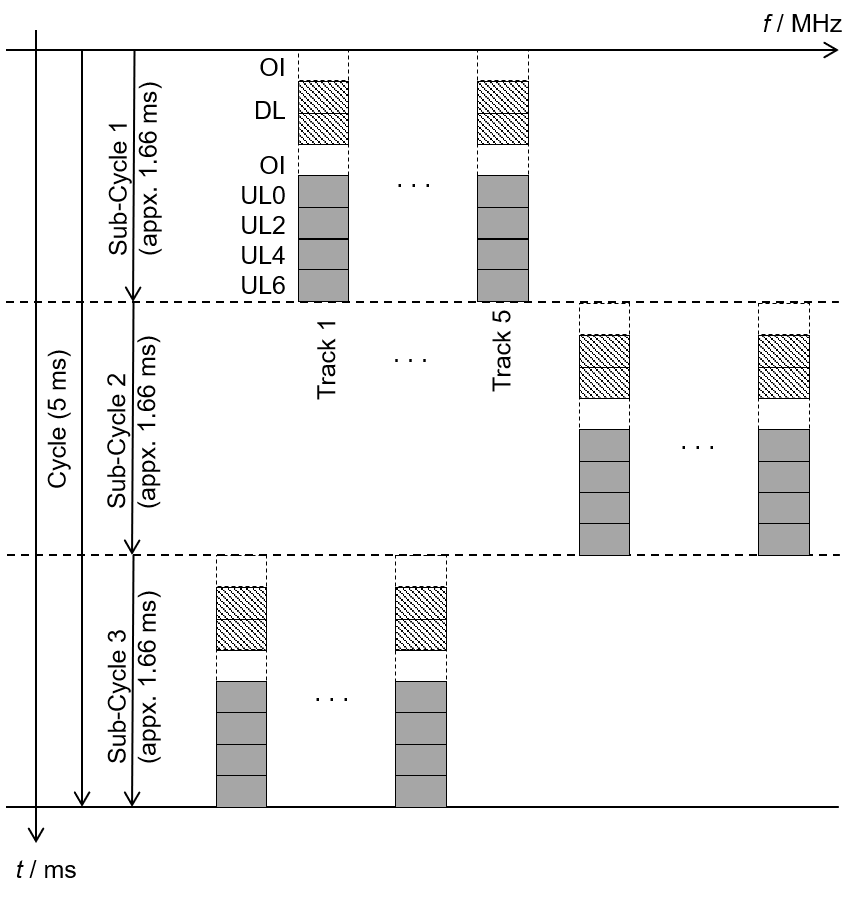}
	\caption{Media access scheme of the IOLW protocol, based on \cite{Heynicke2018}.}
	\label{fig:IOLW-Protocol}
\end{figure}

Compared to other wireless protocols, IOLW provides deterministic media access. Fig. ~\ref{fig:IOLW-Protocol} shows that the communication is divided into cycles and sub-cycles.
A cycle lasts 5\,ms and contains at least three sub-cycles, each of which lasts 1.664\,ms.

Between the sub-cycles, frequency hopping takes place with a hopping distance greater than the typical coherence bandwidth of radio channels in industrial environments  to increase robustness \cite{coexistence}.
% sind diese details wichtig fuer das paper?
% In the organization interval (OI), a switchover between transmitting and receiving, or vice versa is implemented.
% Five tracks with simultaneous downlinks (DL) followed by an OI and four uplink slots (UL) each (numbered evenly starting from zero) are shown, which relates to the maximum configuration level of a W-Master for pure DSlots.

In case of a communication error, a repetition is triggered automatically. With one initial transmission trial and up to two repetitions, a cycle time of $5\,\textrm{ms}$ can be achieved. Due to the specific design of the system and a sufficient link budget, this time is ensured with a remaining failure probability that the cycle time is exceeded of $10^{-9}$ \cite{iolw, Wolberg2018}. Therefore, the average receiving power is presumed to be sufficiently high and the system is not interfered. However, IOLW is not suitable for safety and/or security applications, yet. In this paper, the conceptual approach to enhanced \ac{IOLW} towards a wireless security communication solution for safety critical systems shall be evaluated and relevant security issues should be analyzed.\\

\section{Methodology}
\label{sec:Methodology}
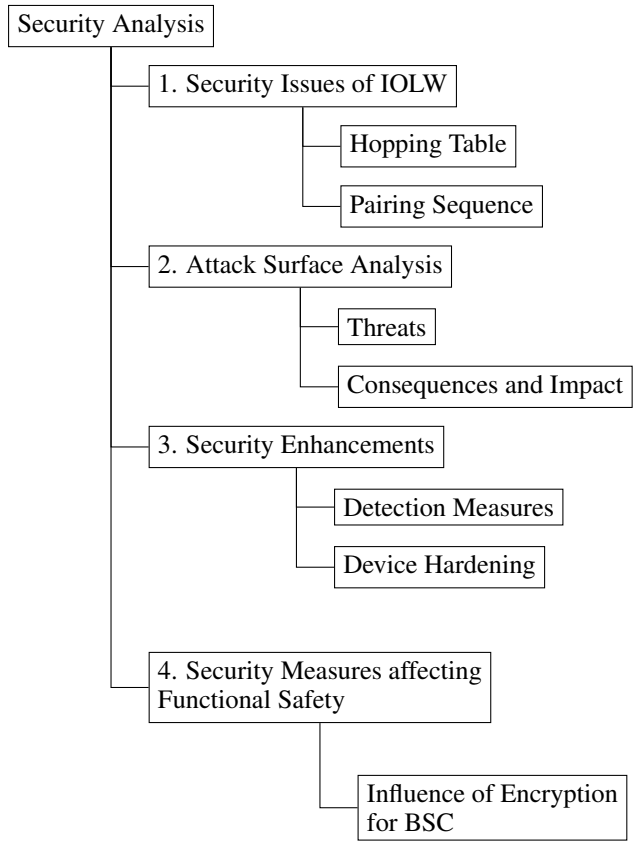
\begin{figure}[t]
\begin{tikzpicture}
[
    %level 1/.style = {red},
    %level 2/.style = {blue},
    %level 3/.style = {teal},
    every node/.append style = {draw, anchor = west},
    grow via three points={one child at (0.5,-0.8) and two children at (0.5,-0.8) and (0.5,-1.6)},
    edge from parent path={(\tikzparentnode\tikzparentanchor) |- (\tikzchildnode\tikzchildanchor)}]

\node {Security Analysis}
    child {node {1. Security Issues of IOLW}
    %child {node [circle, fill, minimum size = 4pt, inner sep = 0] {}}
    %child {node [circle, fill, minimum size = 4pt, inner sep = 0] {}}
	child {node {Hopping Table}}
%    child {node {great-grandchild}}}
	child {node {Pairing Sequence}}
%	child {node {...}}
    edge from parent [solid]}
    child [missing] {}
    child [missing] {}
%    child [missing] {}
    child {node {2. Attack Surface Analysis}
    child {node {Threats}}
%    child {node {great-grandchild}}}
%    child [missing] {}
    child {node {Consequences and Impact}}
    edge from parent [solid]}
    child [missing] {}
    child [missing] {}
    child {node {3. Security Enhancements}
    child {node {Detection Measures}}
    child {node {Device Hardening}}
    edge from parent [solid]}
    child [missing] {}
    child [missing] {}
    child [missing] {}
    child {node [draw, align = left]{4. Security Measures affecting \\ Functional Safety}
    child [missing] {}
    child {node [draw, align = left]{Influence of Encryption \\ for BSC}}
%    edge from parent [solid]}
    edge from parent node [draw = none, left] {}};

\end{tikzpicture}
\caption{Methodology of Security Analysis}
\label{fig:Methodology-SecAnalysis}
\end{figure}

The used methodology depicted in Fig. \ref{fig:Methodology-SecAnalysis}  starts with identifying potential security issues of \ac{IOLW} using the  \ac{IOLW} Systems Extensions - Specification \cite{iolw}.
In the next step, security measures proposed in \cite{doebbert_cammin_scholl_2022} for safety related IOLW applications are identified and in a further step security enhancements are evaluated regarding prerequisites, consequences, safety and security impact and additional mitigation of the specific attack.
Finally, security measures affecting functional safety applications are briefly discussed because this impact is critical and shall be kept in mind. This methodology is then utilized in the following Sections \ref{sec:SecAnalysis} and \ref{sec:InfluenceSecuritySafety}.

%\newpage

\begin{comment}
%TODO TD: Security enhancements beyond [1] --> see Table 1
%TODO TD: Security measures affecting functional safety --> nach weiterer Bearbeitung
\begin{figure}[h!]
	\centering
	\includegraphics[width = 0.48\textwidth]{Material/methodology_flow3.png}
	\caption{Methodology of security analysis.}
	\label{fig:Methodology-SecAnalysis}
\end{figure}
\end{comment}

\vspace*{0.5cm}
% Main part
\section{Security Analysis}
\label{sec:SecAnalysis}
%TODO TD: description of methodology
In this section, an analysis of potential attack surfaces of \ac{IOLW} communication is presented including a threat and risk analysis.

\subsection{Security Issues of IOLW}
\label{subsec:SecIssuesOfIOLW}
%TODO TD: beschreiben
Within this section known security issues and weaknesses of \ac{IOLW} protocol design are outlined.
Therefore, the design of the frequency hopping table and the pairing sequence are described.
Next, attacks on cryptographic measures are described in detail. \\

\subsubsection{Frequency Hopping Table}
IOLW defines dedicated channel hopping sequence algorithms to compute the frequency hopping tables for a W-Master and its W-Devices. These channel hopping sequence algorithms depend on the individual W-Master ID to achieve wireless coexistence within neighboring IOLW systems.
Additionally, a blocklisting can be utilized to avoid certain frequency channels in the computed hopping table favoring wireless coexistence with other wireless systems nearby. \\
Blocklisting is a mechanism to avoid on air collision with other wireless systems, such as WLAN. Conventional Bluetooth cannot be blocklisted, since it is an uncoordinated frequency hopper. The blocklist itself uses eighty 1 MHz wide frequency channels and the master configuration setting can be used to suspend frequency ranges \cite[p.~35]{iolw}.

The default channel hopping table of IOLW is $HT01$ omitting frequencies $f_{1-2}$ and $f_{79-80}$. For configuration, the frequency channels $f_{1}$ and $f_{80}$ are used and $f_{2}$ and $f_{78}$ are Guard-Channels. Blocklisting of each 1\,MHz frequency channel is possible. The carrier frequencies fn in IOLW are defined according
to [3] as $fn = f0 + n  1\,MHz;$ (1) with $f_{0} = 2400MHz$ and $n = 3$ to $78$ (i.e. up to 76 frequency channels) for cyclic data communication. Adaptive Hopping Table - mechanism enables a change of the hopping table of a track while the communication is already running, which may be an improvement of the connection \cite[p.~300-302]{iolw}.\\

\subsubsection{Pairing Sequence}
\ac{IOLW} offers different possibilities to pair a W-Device to a W-Master \cite{iolw}:
\begin{itemize}
\item Pairing by UniqueID, which enables the pairing of an unpaired W-Device to a W-Master Port using a pairing request.
\item Pairing by Button / Re-Pairing can be used to change a damaged W-Device without using a port and device configuration tool (PDCT) or to pair a W-Device to an unused, pre-configured W-Port during  commissioning phase. Therefore, a pairing button or a similar trigger is mandatory for a W-Device.
\item Roaming describes a feature to pair a W-Device temporary to a W-Master, which allows predefined W-Devices between multiple W-Masters.
\end{itemize}

If one W-Master track is in ServiceMode, the configuration channels are available and only in this mode scan, pairing, and roaming activities are possible \cite{iolw}. But in all cases, messages are sent without encryption and authentication. Also, no shared secret between two devices initially exists or is created, which can be used for initial authentication or as temporary authentication or encryption key. The transfer of a W-Device identification in plaintext is fairly insecure.
% IOLW Spec V16 pg. 73 Ch. 5.7 Physical Layer PL protocol

For a security-enhanced pairing sequence, a key establishment technique ideally approved by \cite{FIPS140-2} and recommendations for cryptographic key generations \cite{NIST-800-133r2} are essential. Key exchange possibilities such as \ac{RSA}, \ac{DH}, \ac{ECDH} based key exchange protocol, \ac{ECMQV}, or \ac{CPKE} are described and evaluated for security-enhanced \ac{IOLW} pairing in \cite{doebbert_cammin_scholl_2022}. Here, an \ac{OOB} commissioning technology using \ac{NFC} is suggested to exchange a shared secret. Note that additional hardware is necessary in this case.\\

% see e.g.: https://crypto.stackexchange.com/questions/40200/birthday-attack-on-aes-ccm-with-32-bit-authentication-tag
\subsubsection{Attacks on Cryptographic Measures}
% TODO explain how this crypto was chosen etc.

In \cite{doebbert_cammin_scholl_2022}, AES-CCM with 32-bit or 64-bit authentication tag is suggested and investigated.
Crafting a valid authentication tag would enable an attacker to bypass the cryptographic authentication measure.
Therefore, the probability of a birthday attack on the authentication tag shall be evaluated. \\
\begin{equation}
Adv < \frac{q_{dec}}{2^{\tau}} + \frac{\sigma^{2}}{2^n}
\end{equation}
with \\
$\tau$: length of the tag in bits \\
$\sigma$: number of encrypted / decrypted blocks \\
$n$: block size of the cipher, in bits (128 for AES) \\
$q_{dec}$: allowed number of decryption queries \\
Adv: the probability that an attacker breaks the associated security definition, which shall less than one in 1,000,000 for each random attempt to succeed or a false acceptance to occur \cite{FIPS140-2}. \\

Additionally, in \cite{FIPS140-2} is also stated that ``for multiple attempts to use authentication mechanism during a one-minute period, the probability shall be less than one in 100,000 that a random attempt will succeed or a false acceptance will occur''.
For each attempt, the value of $q_{dec}$ shall be ideally $1$, respectively as low as possible.\\

%TODO multiple IOLW DSlots
For one IOLW DSlot payload (14 bytes) with a \ac{MAC} of $\tau = 32$ using all three uplink messages (message from a W-Device to a W-Master), the following parameter can be assumed:
\begin{equation}
\begin{aligned}
Adv[\tau = 32] & \leq \frac{3}{2^{32}} + \frac{1^{2}}{2^{128}} \\[10pt]
Adv[\tau = 32] & \leq 7 \cdot 10^{-10}
\end{aligned}
\end{equation}

For $\tau = 64$, the probability of crafting a valid MAC is:

\begin{equation}
\begin{aligned}
Adv[\tau = 64] & \leq \frac{3}{2^{64}} + \frac{1^{2}}{2^{128}} \\[10pt]
Adv[\tau = 64] & \leq ~1.6 \cdot 10^{-19}
\end{aligned}
\end{equation}

The lowest transmission time for one message in \ac{IOLW} is about 1.664\,ms and with two retries about 5\,ms. If within two retries no correct message in combination with a valid \ac{MAC} was sent, the W-Master rejects further messages of this W-Device and W-Port. A reconfiguration is needed, therefore the probability values are already based on 1\,min as referred to in \cite{FIPS140-2}.

% bzgl. Sigma: FF: aes-gcm -> 128bit block size -> 16byte block size: 1 block (payload <16byte) -> sigma = 1

\begin{comment}
Therefore, the probability values need to be based on 1\,min:
\begin{equation}
Adv[\tau = 32 \text{, within 1\,min}] > 12,000*
\end{equation}
For $\tau = 64$, the probability of crafting a valid MAC is:
\begin{equation}
Adv[\tau = 64 \text{, within 1\,min}] > 12,000*
\end{equation}
\end{comment}
In both cases, the probability is much smaller than one in 1,000,000 that a random attempt will succeed and, hence the attack to craft a valid MAC is very unlikely. Possibly a shorter \ac{MAC} would also be secure, since a three byte \ac{MAC} results in a probability of about $10^{-7}$ (considering that within one minute, the probability shall be less than one in 100,000). With a two byte \ac{MAC} this consideration of \cite{FIPS140-2} cannot be achieved:

\begin{equation}
\begin{aligned}
Adv[\tau = 16] \leq 4.6 \cdot 10^{-5}
\end{aligned}
\end{equation}

If accepting more faulty packets and the allowed number of decryption queries increases by three tries (including each two retries), the availability of the application increases and the probability that an attacker breaks the associated security definition is still low: %($ADV[\tau = 32, q_{dec}= 10] \leq 7*10^{-9}$).

\begin{equation}
\begin{aligned}
Adv[\tau = 32, q_{dec}= 10] & \leq \frac{10}{2^{32}} + \frac{1^{2}}{2^{128}} \\[10pt]
Adv[\tau = 32, q_{dec}= 10] & \leq 7 \cdot 10^{-9}
\end{aligned}
\end{equation}

Increasing the payload (multiple DSlots), while only using a four byte MAC for e.g. ten DSlots, does not significantly increase the probability that an attacker breaks the associated security definition and is therefore regarded to be sufficient.
Here, the approximation that one DSlot is equal to one encrypted/decrypted block is used. %($ADV[\tau = 32, \sigma= 10] \leq 7*10^{-10}$).

\begin{equation}
\begin{aligned}
Adv[\tau = 32, \sigma= 10] & \leq \frac{3}{2^{32}} + \frac{10^{2}}{2^{128}} \\[10pt]
Adv[\tau = 32, \sigma= 10] & \leq 7 \cdot 10^{-10}
\end{aligned}
\end{equation}

This evaluation depicts that the proposed authentication length is sufficient secure for the probability of an attacker breaking the associated security definition \cite{FIPS140-2}.

% https://nvlpubs.nist.gov/nistpubs/SpecialPublications/NIST.SP.800-38Gr1-draft.pdf
%TODO https://www.ccs.neu.edu/home/wichs/class/crypto-fall17/lecture21.pdf
%"IND-CCA2 security" (security against adaptive chosen plaintext/ciphertext attack).

\subsection{Attack Surface Analysis}
\label{subsec:AttackSurfaceAnalysis}
% Risk scenarios:
This section examines the attack surface for \ac{IOLW} communication using relevant attack scenarios.
Previous mentioned security issues are considered during the creation of attack scenarios.

Protocol specific measures without security background increase the effort for successful attacks drastically.
For instance timeliness of communication bursts and knowledge of the frequency hopping table are two additional hurdles to take for executing a successful attack on the wireless communication.

The consequences of different attack scenarios differ and therefore the overall impact must be investigated in detail.
Therefore within this analysis section the impact for each attack scenario is rated from a security and a safety perspective.
This analysis is used to select and derive suitable protection mechanisms and to rate the effectiveness of already proposed measures.

Security goals within a threat analysis are a common approach to describe the impact of cyber attacks.
A very common security goal model is the CIA-triad, which contains the security goals \textit{Confidentiality}, \textit{Integrity} and \textit{Availability}.
This concept has its origin in the \ac{IT} domain, but is also usable in the \ac{OT} domain.
For the use within this analysis, this model is assumed to be sufficient.
The most important distinction for the \ac{OT} environment is the fact, that availability is the most important goal.

While the use of CIA-triad is common practice within security domain, there exist approaches, where this is also mapped onto safety functionality \cite{Wieczorek}.

A safety function relying on \ac{IOLW} may have two states according to intended disturbance attempts from an attacker - affected or unaffected.
No advantage of a finer distinction of impact from this safety perspective is recognized, therefore only a differentiation of impact or no impact is done.
The result of this attack surface analysis is depicted in Table \ref{tbl:iolw-table-attacks}.

\begin{table*}[h!]
\caption{Possible IOLW attacks influencing the safety and security of its application}
\label{tbl:iolw-table-attacks}
\resizebox{16.5cm}{!}{
\begin{tabular}{@{}|l|l|l|l|l|l|@{}}
\toprule
\multicolumn{1}{|c|}{Attacks}                                                          & \multicolumn{1}{c|}{Prerequisites}                                                                                                                                                                    & \multicolumn{1}{c|}{Consequences}                                                                                                                                                           & \multicolumn{1}{c|}{\begin{tabular}[c]{@{}c@{}}Safety \\ Impact\end{tabular}} & \multicolumn{1}{c|}{\begin{tabular}[c]{@{}c@{}}Security Impact \\ (CIA-triade)\end{tabular}}            & \multicolumn{1}{c|}{\begin{tabular}[c]{@{}c@{}}Additional \\ Mitigation\end{tabular}}                                                \\ \midrule
\begin{tabular}[c]{@{}l@{}}Denial-of-Service \\ (Flooding)\end{tabular}                & \begin{tabular}[c]{@{}l@{}} - Hopping Table,\\ - Knowledge of \\IOLW config\\ - IOLW W-Master in \\ pairing mode\end{tabular} & \begin{tabular}[c]{@{}l@{}}- W-Master is blocked by \\ processing flooding packets\\ - Valid packets cannot be \\ processed in time\\ - W-Master goes back to initial \\ safe mode\end{tabular} & No                                                                            & Availability                                                                                            & \begin{tabular}[c]{@{}l@{}}- Early Attack detection \\ using a IOLW Sniffer\\ - Secure transmission \\ of Hopping Table\end{tabular} \\ \midrule
Jamming                                                                                & \begin{tabular}[c]{@{}l@{}} - Jamming W-Device for all \\ frequencies\end{tabular}                                                                                           & \begin{tabular}[c]{@{}l@{}}- W-Master is blocked by \\ processing flooding packets\\ - Valid packets cannot be \\ processed in time\\ - W-Master goes back to initial \\ safe mode\end{tabular} & No                                                                            & Availability                                                                                            &   Jammer detection                                                                                                                                    \\ \midrule
Replay Attack                                                                          & \begin{tabular}[c]{@{}l@{}}- Hopping Table \\ - Sniff valid traffic\\ - Replay sniffed packets\end{tabular}                                                                                         & \begin{tabular}[c]{@{}l@{}}- W-Master is blocked by \\ processing flooding packets\\ - Valid packets cannot be \\ processed in time\\ - W-Master goes back to initial \\ safe mode\end{tabular} & No                                                                            & Availability                                                                                            &  \begin{tabular}[c]{@{}l@{}}  Replay attack\\ discovery on \\ \ac{IOLW} W-Master \end{tabular}                                                                                                                                   \\ \midrule
Packet Forgery        & \begin{tabular}[c]{@{}l@{}} - Current counter value \\ - Hopping Table     \end{tabular}                                                                                                         & \begin{tabular}[c]{@{}l@{}}- Rare chances to craft valid \\ MAC and valid payload\end{tabular}                                                                                              & Yes                                                                           & \begin{tabular}[c]{@{}l@{}}Availability,\\Integrity\end{tabular} &  \begin{tabular}[c]{@{}l@{}}  Detection of \\malicious packets\\on \ac{IOLW} W-Master   \end{tabular}                                                                                                                               \\ \midrule
\begin{tabular}[c]{@{}l@{}}Packet Forgery \\ (Leaked Key)\end{tabular}                 & \begin{tabular}[c]{@{}l@{}}- Current counter value \\ - Hopping Table \\ - Leaked key \end{tabular}                                                                                                                    & \begin{tabular}[c]{@{}l@{}}- Attacker can create valid \\ MAC for any payload.\end{tabular}                                                                                                 & Yes                                                                           & \begin{tabular}[c]{@{}l@{}}Availability,\\Integrity \end{tabular} & \begin{tabular}[c]{@{}l@{}}IOLW W-Device \\ hardening measures \\ e.g. tamper detection,\end{tabular}                                   \\ \midrule
\begin{tabular}[c]{@{}l@{}}Attacker controls \\ IOLW W-Device \\ (backdoor)\end{tabular} & \begin{tabular}[c]{@{}l@{}}- Physical or remote access to \\ IOLW W-Device \\e.g. via backdoor\end{tabular}
& \begin{tabular}[c]{@{}l@{}}  Full control to create valid \\ \ac{IOLW} messages \end{tabular}  &  Yes                                                                             & \begin{tabular}[c]{@{}l@{}} Availability,\\Integrity,\\ Confidentiality    \end{tabular}                                                                                                    &  \begin{tabular}[c]{@{}l@{}} Device hardening,\\ IDS Systems, \\Anomaly detection in\\ \ac{IOLW} W-Master \end{tabular}                                                                                                                                    \\ \bottomrule
\end{tabular}
}
\end{table*}

Each attack scenario is described with the prerequisites necessary for successful execution of each threat.
Next the consequences of an successful attack are described.
These consequences are then rated for its safety and security impact on a generic use case based on \ac{IOLW} communication.
At the end of each attack scenario description additional mitigations are listed, which are intended to protect against the mentioned threat.
These mitigations are outlined further within the following section.

% Summary of table
The overall outcome of the attack surface analysis, presented in Table \ref{tbl:iolw-table-attacks} is now discussed.
A first noteworthy finding is, that attacks, which only disturb the communication or the timely arrival of messages, are considered having no safety impact.
This is backed by the fact, that common safety peers enter a safe state in such events, where no harm originates from the system.
These events only lead to impacting the availability of the system and are relevant for security observations.

Critical and not considered within safety analysis is the crafting of valid messages and injecting these into the communication system.
Therefore safety functions relying on \ac{IOLW} communication can be manipulated, e.g. safety sensor values become falsified.
These attacks are considered having a severe safety impact.

From a security perspective all considered attack scenarios impact the protection goal availability of the communication system, which is assumed the most targeted goal for attackers.
Disturbance of availability leads directly to financial consequences for the asset owner.
Attacks like flooding, jamming or replay attack are considered easier in its execution, in comparison to the attack scenarios, which also impact the safety.
An attacker with the goal to disturb the availability is assumed to use these attacks instead attacks with more pre-work necessary.

When security enhancements are enabled within the protocol, like proposed in \cite{doebbert_cammin_scholl_2022}, all safety impacting attacks require the bypassing of cryptographic measures, e.g. key knowledge or weak cryptographic measures.
These measures are considered to address the threat of safety impact by state of the art cryptography.

But further mitigations are necessary to detect and react early on attack attempts.
These are presented in the following section.

\subsection{Further Security Enhancements}
\label{subsec:SecEnhancementsBeyond}
%TODO TD

All attack scenarios on \ac{IOLW} communication require physical proximity to the devices.
Therefore, the feasibility of such attacks is generally rated potentially low.
Nonetheless, further measures beyond physical access restrictions are considered necessary, since wireless networks may not be physical protect-able for all use cases.

The following enhanced security implementations are described in the proposed IOLW Safety architecture of \cite{doebbert_cammin_scholl_2022}:
\begin{itemize}
\item[1.] Assessment of the automation environment
\item[2.] Establishing a one-to-one connection
\item[3.] Cryptographic algorithms for IOLW message exchange
\end{itemize}

The assessment of the automation environment includes the handling of pairing and bonding, security parameter negotiation, encryption, key generation and distribution, and e.g. communication to a \ac{HSM}. \\
Establishing a one-to-one connection involves security during commissioning of secrets and security after commissioning of the network key.\\
In the last part, possible cryptographic algorithms for IOLW message exchange, such as AES-CCM and others e.g. in \cite{IOLWcrypto}, are evaluated.

Next to these measures, we derived further mitigation from the listed attack scenarios.
% TODO considered improvements derived from analysis part.
% Mitigations  ################
% general
Attacks on \ac{IOLW} communication itself have multiple preliminaries.
Execution of attacks with real impacts to the communication system are considered difficult, especially due to the need of physical proximity, knowledge of cryptographic key, frequency hoping table and correct time of message exchange.

Further mitigations can be categorized in the following two types, which are depicted now. \\

%TODO add sme more content towards IOLW
\subsubsection{Detection Measures}

% sniffing freq. table
A potential security loophole was identified within the beginning of communication via \ac{IOLW}.
As the hopping table is transmitted in clear text without using any cryptographic measures between the W-Master and W-Device, sniffing of this table is a potential threat.
Possession of the frequency hopping table enables certain previous presented attack scenarios.

A cryptographic protected transmission of the table is seen as a possible solution to address this threat.
If an security-enhanced pairing exchanges a session or link key to encrypt further communication packets, the frequency hopping table could be adjusted post initial pairing.
Another possible solution to protect the transmission of the table can be the transferred via OOB e.g. NFC.

% Detection of invalid/malformed MICs
The integrity and confidentiality is addressed by the use of \ac{MAC}, like proposed in \cite{doebbert_cammin_scholl_2022}.
The limited message length conditions the use of only four bytes or even three bytes for the \ac{MAC}.
The use of truncated \ac{MAC} authenticator length weakens the state of the art cryptography and does not match recommendations for length.
To inhibit brute force attacks, the W-Master should react on receiving wrong \ac{MAC} authenticators with the transition of the communication system towards a fail-state.
The receiving of multiple wrong \ac{MAC} indicates malicious actions ongoing on the involved W-Device.
A shut down of the communication path after detection of irregular behavior reduces the probability of a successful guess for a valid \ac{MAC}.
This behavior results in availability impact of the system, but reduces the potential successful guessing and therefore safety impact.
To sum up these observations, the truncated \ac{MAC} is seen as sufficient measure.

% Sniffer / communcation observation/ Intrusion Detection System (IDS)
To detect and react early on attack attempts further measures should be implemented within the communication environment.
For instance an independent sniffer node could be installed, to detect malicious traffic at an early state of attack.
Especially flooding and jamming attempts are addressed by this measure.
Since no measures within the protocol design can mitigate flooding or jamming attacks, an early detection of such attacks is an essential building block within an overall IOLW security concept. \\

% detections TODO FF
% IOLW Sniffer
% Jammer detection
% Replay attack detection on master
% Malicious packet detection
% IDS/Anomaly detection
% Device hardening TODO FF

\subsubsection{Device Hardering}

% master iolw hardening
Attacks on the IOLW communication are overall rated rather unlikely, since attacks on remotely accessible devices, e.g. the \ac{IOLW} master are more likely and the impact is assumed higher.
Therefore further device hardening measures on the W-Master device should be investigated in future work.

But also \ac{IOLW} W-Device should be hardended by implementing a defense-in-depth protection strategy.
This addresses the attack scenario of fully controlled \ac{IOLW} W-Device for instance by applying secure boot, tamper detection and cryptographic protection for key information stored on the W-Device.

The development cycle should therefore be aligned to the IEC\,62443-4-1 security norm and security measures from IEC\,62443-4-2 should be defined as targeted security levels \cite{IEC62443src}.

\section{Security Measures Affecting Functional Safety}
\label{sec:InfluenceSecuritySafety}

A functional safety protocol is not initially secure and also vice versa a security protocol is not in general safe, because safety mechanisms detect residual errors whereas security mechanisms try to detect errors/manipulations injected purposely or created unintentionally.

The assumption of a certain residual error probability, such as the probability of undetected errors, with a distribution for the black channel principle is typically based on the \ac{BSC} model , e.g. in \cite{IEC61784-3}. Several error types might not be described sufficiently using the \ac{BSC} model, particularly when using security algorithms in the underlying communication layers \cite{SchillerEnhancementSafetyComm}. Security algorithms being \ac{BSC}-preservative and even being viewed as part of the black channel are offering greater efficiency and flexibility \cite{Schiller2020, HorchHannen2019}. Both papers also describe that fail-safe communication mainly focuses on random errors, but lacks cryptographic techniques protecting against attacks.

A particular problem evolves because some well-established assumptions for functional safe mechanisms may become incorrect using cryptographic algorithms. Therefore, an authenticated and encrypted message may become secure, but the necessary risk reduction for a certain \ac{SIL} classification cannot be guaranteed any longer. In this case, functional safety and (cyber-)security are not independently from each other.
This case and complexity is described in detail in \cite{horch2019verschlusselung} with a short safety telegram being transferred encrypted and received with a complete different plaintext as initially intended. The logical consequence is that a sustainable safety protocol must meet modern communication requirements including encryption and the worst possible \ac{BEP} of 0.5 being stated in EN 50159.
%TODO TD: Check again 0.5 BEP
% see: https://webstore.iec.ch/preview/info_iectr63069%7Bed1.0%7Den.pdf
Starting in 2019, a IEC working group (project IEC TR 63069 Ed1) created a framework for functional safety and security, which involves guidance on the common application of IEC 61508 \cite{IEC61508} and IEC 62443 \cite{IEC62443src} in the area of industrial-process measurement, control and automation. The aim of the working group and its technical report is to provide guidance that both fields have no negative influence on each other.

% BSC stuff and probabilities
% see Enhancement_of_safety_comm_model_10.1515_auto-2021-0098.pdf
% maybe black channel principle IEC 61784-3

%================================================================================================
\section{Conclusion}
\label{sec:conclusion}

%================================================================================================
%TODO TD: Summary IOLW und Methodology

Need for security within \ac{IOLW} has been outlined within this contribution using a methodology to identify potential security issues of \ac{IOLW}, evaluating the security measures proposed in a former publication for safety related IOLW applications, and depicting in a further step security enhancements regarding prerequisites, consequences, safety and security impact and additional mitigation of a specific attack.

%Our analysis investigated issues with the current protocol specification, depicted the security measures mentioned within a recent proposal and listed noteworthy cyber attacks on the communication.
%TODO TD: MAC propability calculations
The probability of a birthday attack on the authentication tag has been evaluated for different tag and message length showing that the proposed tag length is sufficient secure.
Also impact on safety and security goals have been rated within this listed evaluation.
The derived security enhancements for the listed attack scenarios have been outlined and can be used for real world applicability evaluation.

The attack scenarios analyzed have all impact on the availability of the communication system.
Safety impact requires bypassing of implemented cryptographic measures.
Attackers with the intention to disturb the availability will more likely execute attacks without safety impact, since attacks like flooding are easier to carry out.

Furthermore, security measures affecting functional safety have been discussed with the influence of creating new models for the residual error probability when applying cryptographic algorithms within the black channel principle. To maintain existing and previous model calculations, a separation of functional safety and security is still eligible.

\section*{Acknowledgment}
%This work has partly been fundet by dtec.bw.
The authors would like to acknowledge Kunbus GmbH, especially D. Krush, D. Krueger, and H. Wattar, and furthermore C. Cammin and R. Heynicke of Helmut-Schmidt-University for their continuous support and rich discussions. \\

\section*{Funding}
This work is funded under the project ``Digital Sensor-2-Cloud Campus Platform'' (DS2CCP) by the Federal Ministry of Defense under the dtec.bw program. Project website: https://dtecbw.de/home/forschung/hsu/projekt-ds2ccp/projekt-ds2ccp

%This publication has been funded by the Open-Access-Fund of the Helmut-Schmidt-University/University of the Federal Armed Forces Hamburg.
% The preferred spelling of the word ``acknowledgment'' in America is without
% an ``e'' after the ``g''. Avoid the stilted expression ``one of us (R. B.
% G.) thanks $\ldots$''. Instead, try ``R. B. G. thanks$\ldots$''. Put sponsor
% acknowledgments in the unnumbered footnote on the first page.

\bibliographystyle{IEEEtran}
\bibliography{literatur}

%\bibliography{bib_all}
%\bibliographystyle{ieeetr}
%\bibliographystyle{IEEEtranC}

\begin{acronym}
  \acro{ICS}{Industrial Control System}
  \acrodefplural{ICS}{Industrial Control Systems}
  \acro{SCADA}{Supervisory Control and Data Acquisition}
  \acro{MAC}{Message Authentication Code}
  \acro{HSM}{Hardware Security Module}
  \acro{IOLW}{IO-Link Wireless}
  \acro{IP}{Internet Protocol}
  \acro{IT}{Information Technology}
  \acro{OT}{Operational Technology}
  \acro{RSA}{Rivest-Shamir-Adleman}
  \acro{DH}{Diffie-Hellman}
  \acro{ECDH}{Elliptic-Curve Diffie-Hellman}
  \acro{ECMQV}{Elliptic-Curve Menezes-Qu-Vanstone}
  \acro{CPKE}{Certificate-based pairwise Key Establishment}
  \acro{OOB}{Out-of-band}
  \acro{NFC}{Near-Field Communication}
  \acro{BEP}{Bit Error Probability}
  \acro{CPS}{Cyber-Physical System}
  \acro{TLS}{Transport Layer Security}
  \acro{SIL}{Safety Integrity Level}
  \acro{BSC}{binary symmetric channel}
  \acro{EUC}{Equipment Under Control}
  \acro{ECC}{Elliptic Curve Cryptography}
\end{acronym}

\end{document}